\documentclass[12pt]{article}

\input epsf
\usepackage{epsfig}
\usepackage{graphicx,epsf,amsmath,amsfonts,amssymb,amsbsy}

\def\d{\mbox{d}}

\title{Non-coherent contributions in charge-exchange
reactions and $\eta$--$\eta'$ mixing }
\author{M. L. Nekrasov \\
{\small\it 
Institute for High Energy Physics, NRC ``Kurchatov
Institute'',} \\
{\small\it Protvino 142281, Russia}  }

\date{}
\begin{document}
\maketitle

\begin{abstract}
We analyse $K^- p \to (\eta, \eta',\pi^0) \Lambda$ on the basis of the
fit of data in a wide region of energies, and $\pi^- p \to (\eta, \eta')
n$ at the energies of GAMS-4$\pi$. We show that disagreements between
the data and the predictions of Regge theory may be explained by the
mode change of summation of intermediate contributions at increasing
energy, from coherent to non-coherent. A method of experimental
measurement of the non-coherent contributions is proposed. On the basis
of available data on the charge-exchange reactions the $\eta$--$\eta'$
mixing is estimated. 
\end{abstract}

\section{Introduction}\label{sec1}

It has been observed \cite{N1} that binary charge-exchange reactions of
hadrons at high energies go through the charge-exchange scattering of
fast quarks located in the beginning of quantum fluctuations, the
splitting and recombination of partons inside hadrons. This fact has
far-reaching effects. First, the mentioned fast quark scattering is
accompanied by high virtualities \cite{N1}. Therefore the appropriate
subprocesses are hard and may be described in the parton model. Second,
the mentioned scattering interrupts the fluctuations, which results in
the formation of a cloud of uncorrelated partons. In the exclusive
reaction they are to be captured by the flying away clusters, and this
induces destruction of the coherence. So one can expect destruction of
the coherence of intermediate contributions. Due to increasing duration
of the interaction with increasing energy of the collisions
\cite{Gribov,Low,Petrov1,Petrov2}, the latter effect should be
increasing with increasing the energy.

Based on the above considerations \cite{N1} proposed a model for the
description of $\pi^- p \to M^0 n$ and $K^- p \to M^0 \Lambda$, $M^0 =
\eta, \eta',\pi^0$. The model also used the idea that contributions of
soft interactions that follow the hard scattering, obey the Regge
behavior \cite{Regge,Irving}. However, the mode of summations of
elementary contributions may be coherent or non-coherent. In the former
case the model reproduced conventional Regge approach. In the case of
non-coherent summation the model gave non-trivial predictions. Since the
mode of summation should change with increasing the energy, one of
non-trivial predictions was the emergence of the energy dependence in
the vertex functions, prohibited in the Regge approach. Such a
dependence is really observed at comparing at different energies the
dependence on the transfer of the ratio of yields of $\eta'$ and $\eta$
in the $\pi^-$ beams \cite{N1,Stanton}. However, the most striking
effect was  disappearance of a dip near $|t| =$ 0.4 (GeV$\!$/c)$^2$ with
increasing the energy in the differential cross-section $K^- p \to
\eta\Lambda$. In reality, the dip is observed at the $K^-$ momentum 3--8
GeV$\!$/c in the laboratory frame \cite{Moscoso,Mason,Marzano,Harran}
and disappears at the momentum 32.5 GeV$\!$/c \cite{GAMS}. In the Regge
approach the dip is explained as a consequence of a dominance of the
vector-exchange trajectory and simultaneously its zeroing by the
signature factor in the region $\alpha_{V}(t) \approx 0$, where
$\alpha_{V}(t)$ is the vector trajectory \cite{Martin}. However, this
mechanism is independent of the energy, and it is not clear why it
ceases to operate with the increasing energy. Ref.~\cite{N1} explained
the effect by the mode change of summation of intermediate
contributions. But the data fit was not carried out in \cite{N1} and
therefore the explanation was a qualitative.

In this paper we carry out a full fit of available data at the
relatively low and high energies, and on this basis we confirm the above
explanation. In addition, we solve the problem of the description at 
intermediate energies where contributions of both modes are possible.
In effect, we propose an algorithm for measuring a relative value of the
coherent and non-coherent contributions. As a by-product we get an
independent estimate of the $\eta$--$\eta'$ mixing. A possible admixture
of the glueball is considered, as well.

We carry out the analysis in a modified version of the model \cite{N1}.
The modification concerns mainly the version of the Regge phenomenology.
First of all, we defreeze the parameters responsible for the quark
symmetry breaking. This modification is particularly important for
determining the $\eta$--$\eta'$ mixing. Further, following \cite{Irving}
we change a parameterization of the vertex functions in the spirit of
Veneziano model. At last, we consider the trajectories non-degenerate
and we take into account the effect of interference between their
contributions.

The paper is organized as follows. In the next section we formulate 
modifications in the model. The data fit is carried out in
sect.~\ref{sec3}. A generalization of the model to a simultaneous
consideration of the coherent and non-coherent contributions is
discussed in sect.~\ref{sec4}. In sect.~\ref{sec5} we discuss the
results and make conclusions. In appendix A we transform data
\cite{GAMS} from the form of numbers of pairs of gamma-quanta to the
form of differential cross-sections.

\section{The model}\label{sec2}
 
As noted in the introduction, the high-energy charge-exchange reactions
occur via the inelastic scattering of fast quarks. The appropriate
subprocesses are hard and can be described in the QCD pertubation
theory. The calculations in the cases of $\pi^- p \to (\eta,\eta',\pi^0)
\, n$ and $K^- p \to (\eta,\eta',\pi^0)\Lambda$ \cite{N1} show that at
low transfers the relevant contributions to the amplitude are equal in
absolute value and independent of the flavors and energies of the
colliding quarks. On the background of hard subprocesses the
contributions of soft processes are further formed. We assume that they
are independent of the flavors in the hard subprocesses. In this case
the contributions of hard subprocesses actually are factorized, and
their non-trivial result is a certain mode of summation of soft
contributions. For a quantitative description of the soft contributions
we apply the Regge phenomenology. In doing so, we consider contributions
associated with the valence quarks in the final state as elementary
ones. They are formed in different ways and they are summed differently
in the different modes. As a result, in the mode of coherent summation
with a certain choice of the signs the formulas of the conventional
Regge approach are reproduced. In the non-coherent mode the elementary
contributions are summed in the cross-section. In general, this leads to
non-trivial consequences.

Basically, the above discussion defines the model up to the definition
of Regge amplitudes. Turning to the latter issue, we recall that binary
processes in the leading approximation at large $s$ and small $t$ are
described as a sum of contributions of the leading trajectories
\cite{Regge,Irving},
\begin{equation}\label{F1}
A_{ab}(s,t) \; = \; \sum_{i} \beta_{aib}^{\pm}(t) \;
\frac{1 \pm e^{-i\pi\alpha_{i}^{\pm}(t)}}
     {\sin \left(\pi\alpha_{i}^{\pm}(t)\right)} \;
\left( s/s_0 \right)^{\alpha_{i}^{\pm}(t)} .
\end{equation}  
Here $a$ and $b$ mean initial and final states, $\alpha_{i}^{\pm}(t)$
are the trajectories of particular parity, $\beta_{aib}^{\pm}(t)$ are 
the vertex functions, $s_0$ is a scale parameter. The numerator in
(\ref{F1}) represents the signature factor. The zeros in the denominator
give Regge poles in the region of bound states ($t>0$). In the
scattering region ($t<0$) the poles must be compensated by zeros in
$\beta_{aib}^{\pm}(t)$. The latter property is explicitly realized in
the parameterization in the spirit of the Veneziano model, which
includes the gamma function instead of the sine in the denominator
\cite{Irving}:
\begin{equation}\label{F2}
A_{ab}(s,t) \; = \; \sum_{i} \beta_{aib}^{\pm}(t) \;
\Gamma\left(1-\alpha_{i}^{\pm}(t)\right) 
\left[1 \pm e^{-i\pi\alpha_{i}^{\pm}(t)}\right]     
\left( s/s_0 \right)^{\alpha_{i}^{\pm}(t)} .
\end{equation} 
Here vertex functions $\beta_{aib}^{\pm}(t)$ differ from those in
(\ref{F1}) and have smoother behavior. Phenomenologically they are well
described by an exponential function, possibly with a modification by
polynomial factors describing spin-flip contributions. 

In fact, the presence of exponential factors in the vertex functions is
set by formula (\ref{F2}) itself. Really, the r.h.s in (2) does not
change with simultaneous  substitutions $s_0 \to \tilde{s}_0$,
$\beta_{aib}(t) \to \tilde{\beta}_{aib}(t)$, where
\begin{equation}\label{F3}
\tilde{\beta}_{aib}(t) = \beta_{aib}(t) 
\left( \tilde{s}_0/s_0 \right)^{\alpha_{i}(t)}.
\end{equation} 
Hereinafter we omit the sign of the signature ($\pm$) assuming it is
included in the index $i$. In linear approximation for the trajectory,
$\alpha_{i}(t) = \alpha_{i}(0) + \alpha'_{i}(0)\,t$, we have
\begin{equation}\label{F4}
\tilde{\beta}_{aib}(t) = 
\left( \tilde{s}_0/s_0 \right)^{\alpha_{i}(0)} \beta_{aib}(t) \times 
e^{\alpha'_{i}(0)\,\ln(\tilde{s}_0/s_0) \, t } \,.
\end{equation} 
So, if at some $s_0$ the vertex functions do not include an exponential
factor, then this factor does appear at the transition to another scale
parameter. For example, in the Veneziano model the vertex
factor initially appears as a constant at the scale parameter $s_0 =
1/\alpha'$, where $\alpha'$ is a slope of the trajectory. However,
without changing the amplitude, we can choose another scale parameter
with simultaneous change of the vertex function in accordance with
(\ref{F4}). In particular, at the transition to the scale $\tilde{s}_0$
the vertex factor in the Veneziano model is converted to an exponential
function with the slope $\alpha' \ln(\tilde{s}_0 \alpha')$. 

Below we assume that the vertex functions have purely exponential
behavior (up to the spin-flip factors) and we consider their
normalizations and slopes as free parameters. Simultaneously we put $s_0
= s$ in formula (\ref{F2}). This condition unifies the normalizations
and collects exponential $t$-dependence completely in the vertex
functions. At the same time, at the transition to different energies the
results can be recalculated by means of (\ref{F4}). 

Further, we assume that the slopes in the vertex functions are
determined by the trajectories, not by real particles in the final
state. So at small $t$, we put 
\begin{equation}\label{F5}
\beta_{aib}(t) = \beta_{aib}(0) \exp( c_i t )\,.
\end{equation}  
The dependence on real particles is included in the norma\-lization
$\beta_{aib}(0)$, which is manifested in the overall strength of the
coupling, in the group factors and the mixing parameters. Violation of
the flavor symmetry is described by additional exponent-like factors,
\begin{equation}\label{F6}
\xi_{i} = \xi_{0i} \; \exp(\xi'_{i} t)\,.
\end{equation}
Here $\xi_{0i}$ and $\xi'_{i}$ are determined by the trajectory and the
valence quarks in the final state. (Indices for the valence quarks are
omitted to avoid bulkiness.) Actually (\ref{F6}) defines the splitting
of the contributions of trajectories due to violation of the flavor
symmetry.\footnote{Here we proceed from the provision, that the
splitting accumulates during formation of the trajectories and is weakly
related to the formation of real particles in the final state. In
support we note that the characteristic time of the former process is
much greater than that of the latter one. Really, the formation time of
a relativistic particle is of order ${\cal E}/\mu^2$, where ${\cal E}$
and $\mu$ are its energy and mass. In the case of trajectories the
characteristic time is determined by the formation time of fast partons
(at the recombination of the fluctuations). Since the masses of the
partons are much smaller than the hadron masses and their momenta are
comparable, the above ratio occurs.}

Let us consider for definiteness the charge-exchange processes in the
$\pi^-$ beams with yields of $\eta$ and $\eta'$. In the leading
approximation they are determined by the $a_2$-trajectory \cite{Regge}
and the corresponding formulas are independent of the mode of summation
of intermediate contributions \cite{N1}. Taking into account above
comments, we proceed directly to the differential cross-sections,
\begin{eqnarray}\label{F7}
 \frac{\d\sigma}{\d t} (\pi^{-} p \to \eta  n)\, &=&
g_{a_2}^2 (t) \,  (1 - r t) \,
 \cos^2 \!\frac{\pi \alpha_{a_2}}{2}
 \left(\cos\theta - \xi_{a_2}\sqrt{2}\sin\theta\right)^2 \,,
\\ [0.2\baselineskip] \label{F8}
 \frac{\d\sigma}{\d t} (\pi^{-} p \to \eta' n) &=&
g_{a_2}^2 (t) \,  (1 - r' t)
 \cos^2 \!\frac{\pi \alpha_{a_2}}{2}
 \left(\sin\theta + \xi_{a_2}\sqrt{2}\cos\theta\right)^2 \,.
\end{eqnarray}
Here in both formulas the first factor is a flavor-indepen\-dent
contribution to the vertex function. In the case of arbitrary trajectory
``$i\,$'', in accordance with (\ref{F2}) and (\ref{F5}), it is defined
as
\begin{equation}\label{F9}
g_{i} (t)  =  
g_{0 i} \; \Gamma\left(1-\alpha_{i}(t)\right) \; \exp(c_{i} t)\,,
\end{equation}
where $g_{0 i}$ and $c_{i}$ are phenomenological parameters. (Remember,
in (\ref{F7}) and (\ref{F8}) $i=a_2$). The second factor in (\ref{F7}),
(\ref{F8}) is the spin-flip contribution. The third is the signature
factor. The last factors stand for flavor-dependent contributions to the
vertex functions. They are determined by the angle of $\eta$--$\eta'$
mixing, by the group factors, and by the nonet-symmetry-breaking factor
$\xi_{a_2}$ introduced in (\ref{F6}). Here, we consider the mixing in
the simplest scheme \cite{PDG},
\begin{eqnarray}\label{F10}
|\eta \rangle\, &=&
 \cos\theta \; |\eta^8 \rangle \, - \,\sin\theta \; |\eta^0 \rangle ,
 \nonumber\\ [0.4\baselineskip]
 |\eta' \rangle &=& \,
 \sin\theta \; |\eta^8 \rangle \, + \,\cos\theta \; |\eta^0 \rangle .
\qquad
\end{eqnarray}
It is worth mentioning that in the case of exact nonet symmetry
($\xi_{a_2} \! = \! 1$) the sine and cosine in (\ref{F7}), (\ref{F8})
define the non-strange component in the wave functions of $\eta$,
$\eta'$. 

By this means formulas (\ref{F7}), (\ref{F8}) describe the reactions
with the aid of seven parameters. Six of them, $g_{0 a_2}$, $c_{a_2}$,
$r'$, $r$, $\xi_{0a_2}$, $\xi'_{a_2}$, are specific. Parameter $\theta$
is universal. 

An important characteristic in the $\pi^-$ beams is the ratio of the
differential cross-sections. After simple transformations, we get
\begin{equation}\label{F11}
 R_{\pi}^{\,\eta'/\eta}(t)    \equiv 
\frac{{\rm d} \sigma / {\rm d} t (\pi^- p \to \eta' n)}
 {{\rm d} \sigma / {\rm d} t (\pi^- p \to \eta  n)}  = 
 \frac{1 \! - \! r' t}{1 \! - \! r t} \tan^2 (\theta + \theta_{id} -
\delta)\,.
\end{equation}
Here $\theta_{id} = \arctan\!\sqrt{2}$ \ ($\theta_{id} \approx
54.7^0$) and
\begin{equation}\label{F12}
\delta = \arctan \frac{\sqrt{2}(1-\xi_{a_2})}{1+2\xi_{a_2}}\,.
\end{equation}
It is seen from (\ref{F11}) that at zero transfer $R_{\pi}^{\eta'/\eta}$
is determined by the difference $\theta - \delta_0$, where $\delta_0 =
\delta(0)$. Traditionally $R_{\pi}^{\eta'/\eta}(0)$ is used for
determining the mixing angle $\theta$. However, we see that on this
basis only the difference $\theta - \delta_0$ can be determined, not 
$\theta$ alone. 

In the case of charge-exchange reactions in the $K^-$ beam with yields
of $\eta$ and $\eta'$, there are two leading trajectories, $K^{*}$ and
$K^{*}_2$ \cite{Regge}. They have different signatures and different
symmetry properties, which complicates the description. Moreover, the
trajectories can be non-degenerate. In this case an interference therm
should appear.

In the coherent mode the model leads to the following formulas for
differential cross-sections (the same formulas arise in the Regge
approach):
\begin{eqnarray}\label{F13}
& \displaystyle \frac{\d\sigma}{\d t} (K^{-} p \to \eta \Lambda) \;=\; 
\qquad &
\nonumber
\\ [0.2\baselineskip]
& \displaystyle 3 \, g_{_{V}}^2 (t) \, \sin^2 \!\frac{\pi
\alpha_{_{V}}}{2} 
\cos^2 \theta \,+\,
\frac{1}{3} \, g_{_{T}}^2 (t) \, \cos^2 \frac{\pi \alpha_{_{T}}}{2}
\left(\cos\theta + 2\sqrt{2}\,\xi\,\sin\theta \right)^2 \qquad &
\nonumber
\\ [0.5\baselineskip]
& \displaystyle  - \; 2 \, g_{_{V}} (t) \, g_{_{T}} (t) \, 
\cos \! \frac{\pi \alpha_{_{T}}}{2} \, 
\sin \! \frac{\pi \alpha_{_{V}}}{2}   \,
\sin \! \frac{\pi (\alpha_{_{V}}\!-\!\alpha_{_{T}})}{2} \;
\cos\theta \left(\cos\theta + 2\sqrt{2}\,\xi\,\sin\theta \right)\,,
\qquad  &
\end{eqnarray}  
\begin{eqnarray}\label{F14}
& \displaystyle \frac{\d\sigma}{\d t} (K^{-} p \to \eta' \Lambda) \;=\;
\qquad &
\nonumber
\\ [0.2\baselineskip]
& \displaystyle 3 \, g_{_{V}}^2 (t) \, \sin^2 \!\frac{\pi
\alpha_{_{V}}}{2} 
\sin^2 \theta \,+\,
\frac{1}{3} \, g_{_{T}}^2 (t) \, \cos^2 \frac{\pi \alpha_{_{T}}}{2}
\left(\sin\theta - 2\sqrt{2}\,\xi\,\cos\theta \right)^2 \qquad &
\nonumber
\\ [0.5\baselineskip]
& \displaystyle  - \; 2 \, g_{_{V}} (t) \, g_{_{T}} (t) \, 
\cos \! \frac{\pi \alpha_{_{T}}}{2} \, 
\sin \! \frac{\pi \alpha_{_{V}}}{2}   \,
\sin \! \frac{\pi (\alpha_{_{V}}\!-\!\alpha_{_{T}})}{2} \;
\sin\theta \left(\sin\theta - 2\sqrt{2}\,\xi\,\cos\theta \right)\,.
\qquad &
\end{eqnarray}
Hereinafter we introduce indices {\small $V$} and {\small $T$} instead
of $K^{*}$ and $K^{*}_2$, respectively, and we omit $T$ in $\xi_{_{T}}$.
The spin-flip factors are not included since the data do not need this
\cite{Moscoso,Mason,Marzano,Harran}. So, above formulas include seven
parameters: $\theta$, $g_{_{V}}$, $c_{_{V}}$, $g_{_{T}}$, $c_{_{T}}$,
$\xi_{0}$, $\xi'$, where $\xi_{0}$ and $\xi'$ are involved in $\xi$,
{\it cf.}~(\ref{F6}). Notice that the vertex functions with purely
singlet final states, in conformity with antisymmetric properties, are
zero in the case of vector trajectories. Accordingly, $\xi$ does not
appear in the vector channel. 

It is helpful noting that expressions in the large round brackets in
(\ref{F13}), (\ref{F14}) are reduced to cosine and sine of $\theta +
\widetilde\theta_{id} - \widetilde\delta$, where $\widetilde\theta_{id}
= -\arctan(2\sqrt{2})$ and
\begin{equation}\label{F15} 
\widetilde\delta = -\arctan \frac{2\sqrt{2}(1-\xi)}{1+8\xi}\,.
\end{equation}
With the aid of this property it is easy to understand the reason of the
appearance of a dip in $K^{-} p \to \eta \Lambda$. The point is that
$(\cos\theta \! + \! 2\sqrt{2}\,\xi\,\sin\theta)$ in the tensor
contributions in (\ref{F13}) is proportional to $\cos(\theta +
\widetilde\theta_{id} - \widetilde\delta)$, which in view of
$\widetilde\theta_{id} \approx -70.5^0$ is approximately zero at $\theta
\approx -20^{\mbox{\scriptsize o}}$, $\widetilde\delta \approx 0$.
Consequently the contributions of the tensor trajectory in (\ref{F13})
are strongly suppressed in the region where $\alpha_{K^{*}}(t) \approx
0$, {\it i.e.} at $t \approx - 0.4$ (GeV$\!$/c)$^2$ \cite{Martin}. 

In the non-coherent mode the formulas for the differential
cross-sections are significantly different. Recall that in this case
the elementary contributions, associated with the valence quark, are
summed in the cross-section. In doing so, the strange and non-strange
quark-antiquark valence pairs appear in the mesonic final states in the
equal parts  (as in the coherent mode, as well) \cite{N1}. This gives
\begin{eqnarray}\label{F16}
& \displaystyle
\frac{\d\sigma}{\d t} (K^{-} p \to \eta \Lambda) \;=\;
\frac{5}{3}\; g_{_{V}}^2(t) \, \sin^2 \!\frac{\pi\alpha_{_{V}}}{2}
                             \, \cos^2 \theta \qquad & \nonumber
\\ [0.2\baselineskip]
& \displaystyle
+ \; \frac{1}{3} \; g_{_{T}}^2(t) \, \cos^2 \frac{\pi
\alpha_{_{T}}}{2}\left[
\left( \cos\theta - \sqrt{2}\,\xi\,\sin\theta \right)^2 +
\left(2\cos\theta + \sqrt{2}\,\xi\,\sin\theta \right)^2 \right] \qquad
& \nonumber
\\ [0.2\baselineskip]
& \displaystyle
+ \; \frac{2}{3} \, g_{_{V}}(t) \, g_{_{T}}(t) \;
\cos \! \frac{\pi \alpha_{_{T}}}{2} \, 
\sin \! \frac{\pi \alpha_{_{V}}}{2}   \,
\sin\frac{\pi (\alpha_{_{V}}\!-\!\alpha_{_{T}})}{2} \;
\cos\theta 
\left(5\cos\theta + \sqrt{2}\,\xi\,\sin\theta \right), \quad &
\end{eqnarray}  
\begin{eqnarray}\label{F17}
& \displaystyle
\frac{\d\sigma}{\d t} (K^{-} p \to \eta'\Lambda) \;=\;
\frac{5}{3}\; g_{_{V}}^2(t) \, \sin^2 \!\frac{\pi\alpha_{_{V}}}{2}
                             \, \sin^2 \theta \qquad & \nonumber
\\ [0.2\baselineskip]
& \displaystyle
+ \; \frac{1}{3} \; g_{_{T}}^2(t) \, \cos^2 \frac{\pi
\alpha_{_{T}}}{2} \left[
\left( \sin\theta + \sqrt{2}\,\xi\,\cos\theta \right)^2 +
\left(2\sin\theta - \sqrt{2}\,\xi\,\cos\theta \right)^2 \right] \qquad
& \nonumber
\\ [0.2\baselineskip]
& \displaystyle
+ \; \frac{2}{3} \, g_{_{V}}(t) \, g_{_{T}}(t) \;
\cos \! \frac{\pi \alpha_{_{T}}}{2} \, 
\sin \! \frac{\pi \alpha_{_{V}}}{2}   \,
\sin\frac{\pi (\alpha_{_{V}}\!-\!\alpha_{_{T}})}{2} \;
\sin\theta 
\left( 5\sin\theta  - \sqrt{2}\,\xi\,\cos\theta \right). \quad &
\end{eqnarray} 
These formulas include the same parameters as (\ref{F13}), (\ref{F14}),
but the parameters values may be different.

Finally, we note that there is a third charge-exchange reaction
in the $K^-$ beams, the $K^{-} p \to \pi^0 \Lambda$, which is determined
by the same trajectories. In this case the formula for the differential
cros-section is independent of mode of summation of elementary
contributions \cite{N1}. Owing to the absence of mesonic singlets in the
final state, it does not contain the mixing parameter $\theta$ and the
singlet-channel splitting $\xi$. However, a similar splitting can occur
because of another isotopic spin in the final state. So we have
\begin{eqnarray}\label{F18}
& \displaystyle
\frac{\d\sigma}{\d t} (K^{-} p \to \pi^0 \Lambda) \;=\; 
\zeta^2 \, \biggl\{
g_{_{V}}^2 (t)  \, \sin^2 \!\frac{\pi \alpha_{_{V}}}{2}  + 
g_{_{T}}^2 (t) \, \cos^2 \frac{\pi \alpha_{_{T}}}{2} \biggr. & \nonumber
\\ [0.5\baselineskip]
&  \biggl. \displaystyle + \; 2 \, g_{_{V}} (t) \, g_{_{T}} (t) \;
\sin \!\frac{\pi \alpha_{_{V}}}{2} \, 
\cos \frac{\pi \alpha_{_{T}}}{2} \,
\sin\frac{\pi (\alpha_{_{V}}\!-\!\alpha_{_{T}})}{2} \biggr\}\,. &
\end{eqnarray} 
Here
\begin{equation}\label{F19}
\zeta = \zeta_{0} \; \exp(\zeta' t),
\end{equation}
and are $\zeta_{0}$, $\zeta'$ are parameters. So, the process with
$\eta$,
$\eta'$, $\pi^0$ are described in total by nine parameters.

\section{The fit}\label{sec3}

To determine the mode of summation of intermediate contributions we
proceed to the fit of data. Recall that in accordance with previous
qualitative analysis \cite{N1} at relatively low and high energies the
coherent and non-coherent mode is realized, respectively. Our task
is to confirm or deny this result on the basis of the fit of data.
Correspondingly, we do the fit of data at relatively low and high
energies independently. In order to eliminate false solutions, we
introduce restrictions on the parameters. Namely, we assume that a
solution is physical if $\theta$, $\xi_0$, $\zeta_0$  belong to
intervals $-35^{\mbox{\scriptsize o}} < \theta < -5^{\mbox{\scriptsize
o}}$, \ $0.5 < \xi_0 <1.5$, $0.5 < \zeta_0 <1.5$. The first condition
cuts off solutions clearly inconsistent with the results of other
studies \cite{PDG}. The second and third conditions mean that the
violation of the flavor symmetry should not be too large. A similar
condition for the slope parameters implies that $\xi'$ and $\zeta'$ in
absolute value should not exceed $c_{_{T}}$ and $c_{_{V}}$, and within
each trajectory the slopes should not split significantly. We demand
also a positivity of the resulting slopes in the vertex functions, which
means decreasing of contributions of the trajectories with increasing
$-t$. Lastly, we consider linear trajectories and we define them based
on the spectroscopy~data.

\subsection{{\boldmath $\pi^-$} beams}\label{sec3.1}

\begin{figure*}
\hspace*{0.08\textwidth}
\includegraphics[width=0.31\textwidth]{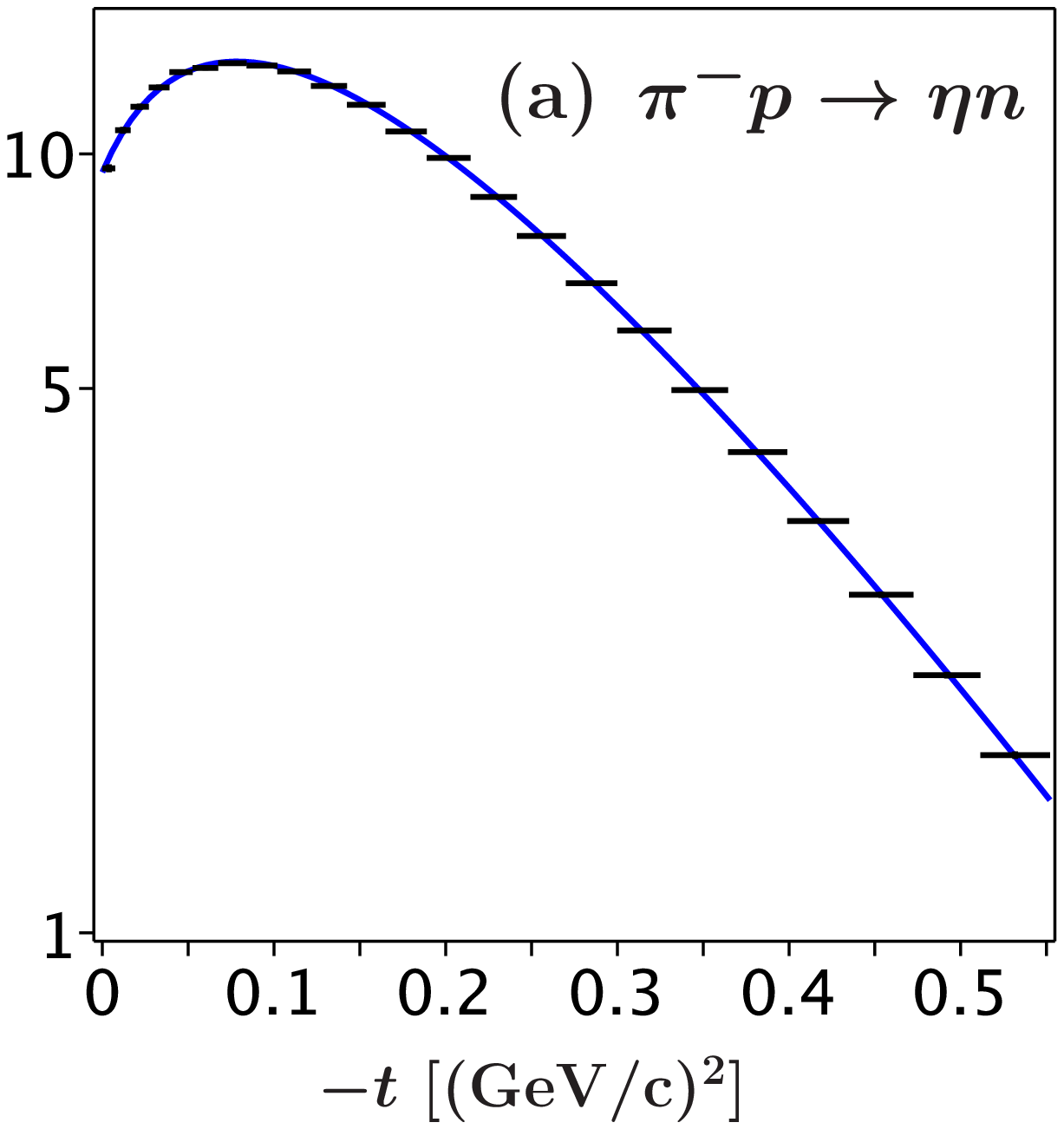}
 \hfill 
 \includegraphics[width=0.31\textwidth]{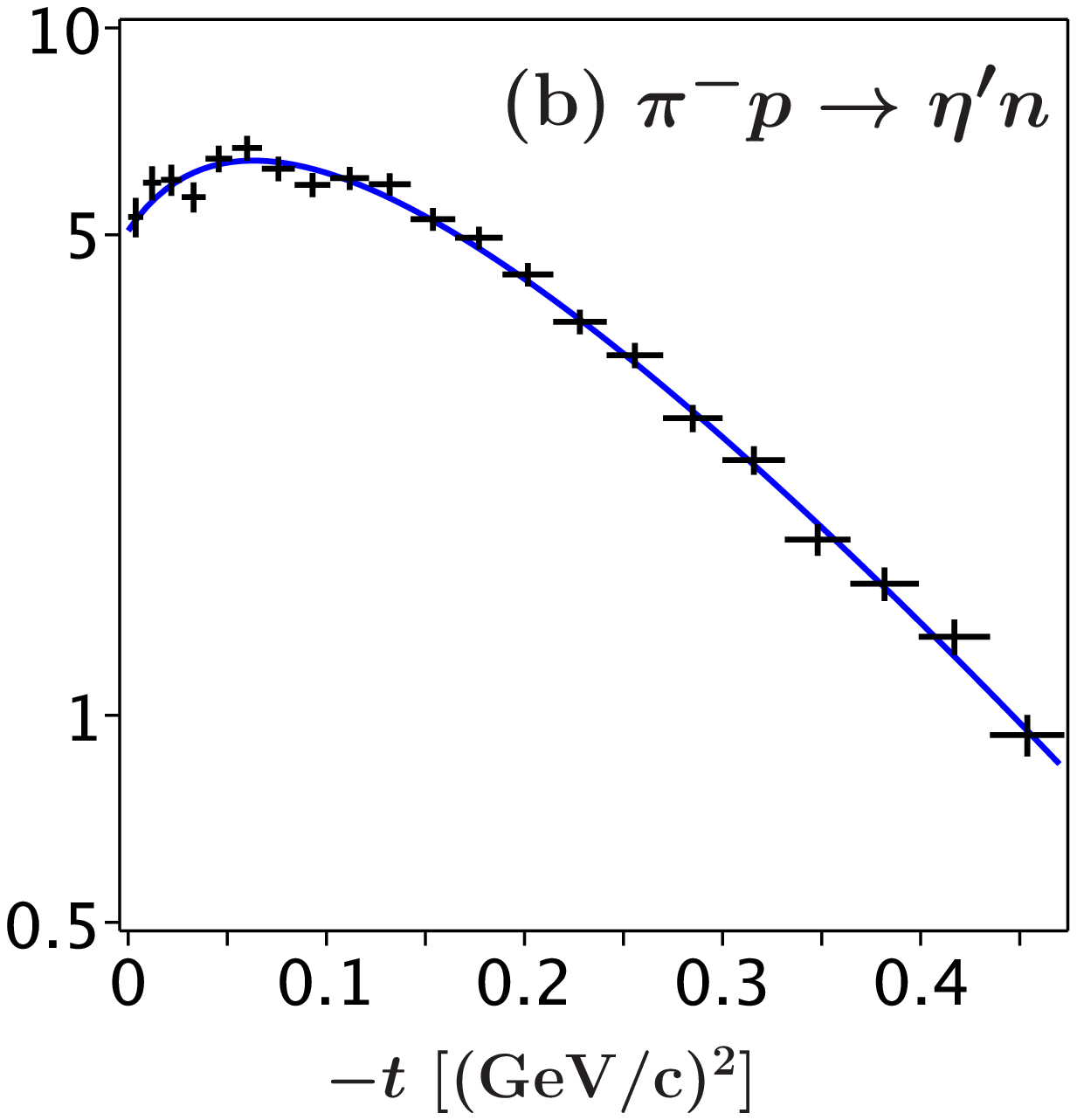}
\hspace*{0.1\textwidth}
\caption{\small Differential cross-sections $\pi^{-} p \to (\eta, \eta')
n$ [$\mu$b/(GeV$\!$/c)$^2$] at 32.5 GeV$\!$/c \cite{GAMS}.
Theoretical curves correspond to (\ref{F7}), (\ref{F8}) with parameter
values specified in the text.}
\label{Fig1} 
\end{figure*}

In the first place we consider data in the $\pi^{-}$ beams. Recall that
appropriate cross-sections are given in (\ref{F7}), (\ref{F8}) and they
are independent of the mode of summation of intermediate contributions.

Based on the spectroscopy data \cite{PDG} the $a_2$-trajectory in the
linear approximation is
\begin{equation}\label{F20}
 \alpha_{a_2}(t) = 0.45 + 0.89\,t \,,
\end{equation} 
where $t$ is given in (GeV$\!$/c)$^2$. The errors in (\ref{F20}) are
of order of percents, which is insignificant for subsequent results.
Notice that earlier determinations of $\alpha_{a_2}(t)$ proceeding
directly from the scattering data, gave $\alpha_{a_2}(t) = 0.4 + 0.7\,t$
\cite{apel} and $\alpha_{a_2}(t) = 0.37 + 0.79\,t + 0.03\,t^2$
\cite{Gahl}. The discrepancy with our result (\ref{F20}) is explained by
the differences in the formulas for the differential cross-sections used
in the mentioned references.

The charge-exchange reactions in the $\pi^{-}$ beams with yields of
$\eta$, $\eta'$ have been studied in detail at 8.45 GeV$\!$/c
\cite{Stanton} and at higher momenta \cite{GAMS,apelp,apel}.
Unfortunately, data \cite{Stanton} and \cite{apelp} are not available,
and data \cite{apel} without \cite{apelp} are not complete. For this
reason we to consider only data \cite{GAMS} at 32.5 GeV$\!$/c. (Of
course, this narrows our capabilities, but we will be able to work out
important details and verify the accuracy of the model itself.) The fit
with these data reveals a series of solutions with close $\chi^2$
distributed in the ($\theta$,$\,\xi_0$)-plane near a curve defined by
condition $\theta - \delta_0 = -18.5^{\mbox{\scriptsize o}}$. When doing
the fit with different fixed ($\theta,\,\xi_0$), the surface of minima
of $\chi^2$, which appears over the ($\theta,\,\xi_0$)-plane, has a
trough of almost constant depth located along the mentioned curve. The
boundaries of the curve are determined by the boundaries of the physical
region for the parameters:
($-5^{\mbox{\scriptsize o}}$, $0.62$) and ($-28.6^{\mbox{\scriptsize
o}}$, $1.5$). In these points $\chi^2/\mbox{d.o.f.}$ takes values
30.4/35 and 30.0/35, respectively, and along the curve $\chi^2$ remains
within the mentioned limits. (The other parameters vary typically within
10--20\%.) So, all the points of the curve determine practically
equivalent solutions. 

For an illustration we point out one of the solutions that appear when
all the parameters are free: $\theta \! = \!\! -21^{\mbox{\scriptsize
o}} \! \pm 8^{\mbox{\scriptsize o}}$, $g_{0 a_2} \! =  1.7 \pm 0.4$,
$c_{a_2} \! = 3.37 \pm 0.08$, $\xi_{0 a_2} \! =  1.1 \pm 0.4$,
$\xi'_{a_2} \! = 0.40 \pm 0.08$, $r \! = \! 19.7 \pm 0.3$, $r' \! = 18.4
\pm 1.7$, $\chi^2/\mbox{d.o.f.} \! = \! 31.0/37$. Hereinafter the
normalization constant(s) are determined in $\mu$b$^{1/2}$/(GeV$\!$/c)
and the slopes in (GeV$\!$/c)$^{-2}$ (see footnote\footnote{We have
recalculated data \cite{GAMS}, obtained initially in the form of numbers
of pairs of gamma-quanta arising from $\eta$ and $\eta'$, into the units
of differential cross-sections, see appendix A. The error of the
conversion factor is not included in $g_{0 a_2}$.}). In fig.~\ref{Fig1}
the corresponding differential cross-sections are presented together
with the data used in the fit.   

It is worth noting that in all the solutions the errors of $\theta$ and
$\xi_{0 a_2}$ are significant. However, both these parameters are highly
correlated and their difference is determined quite accurately. In
particular, for the solution given above $\theta - \delta_0 = (-18.5
\pm 0.6)^{\mbox{\scriptsize o}}$, and this result is kept for all the
solutions along the curve. From here and (\ref{F11}) we restore the
ratio of the cross sections $R_{\pi}^{\eta'/\eta}(0) = 0.53 \pm 0.02$
with $R_{\pi}^{\eta'/\eta}(0) = 0.54 \pm 0.04$ in \cite{GAMS}. In fact,
$R_{\pi}^{\eta'/\eta}(0)$ is practically independent of energy. Really,
\cite{Stanton} obtained $R_{\pi}^{\eta'/\eta}(0) = 0.500 \pm 0.092$ at
8.45 GeV$\!$/c, and \cite{apelp} obtained $R_{\pi}^{\eta'/\eta}(0) =
0.55 \pm 0.06$ at 25 GeV$\!$/c and 40 GeV$\!$/c. So, regardless
$\theta$, the value of $\xi_{0 a_2}$ remains constant (within errors)
with changing the energy. This means that $\xi_{0 a_2}$ is independent
of (weakly dependent on) the mode of summation of intermediate
contributions. However, $\xi'_{a_2}$ depends significantly which follows
from the energy dependence of the slope of $R_{\pi}^{\,\eta'/\eta}(t)$,
see (\ref{F11}), (\ref{F12}) and discussion in \cite{N1}.

Concluding, we note that high reliability of the solutions along the
curve means that our model describes data well and it is quite suitable
for the analysis of the charge-exchange reactions. In particular, our
ansatz about the presence of the gamma function in the amplitude and
the parameterization of the vertex functions is confirmed. If one
removes the gamma function but preserves the signature factor, the
quality of the description falls catastrophically: in the physical
region the minimum of $\chi^2/\mbox{d.o.f.}$ constitutes $126/35$.

\subsection{{\boldmath $K^-$} beams}\label{sec3.2}

{$\!$\it A priori}, we do not know whether the leading $K^{*}$
and~$K^{*}_2$ trajectories are non-degenerate. An independent fit of
spectroscopy data \cite{PDG} clearly indicates a preference of their
non-degeneracy.\footnote{The hypothesis of the degeneracy leads to
$\chi^2/\mbox{d.o.f.} = 26.6$ against $\chi^2/\mbox{d.o.f.} = 1.7$ in
the case of non-degeneracy.} Assuming linearity of the trajectories, we
obtain 
\begin{equation}\label{F21}
\alpha_{K^{*}}   = 0.33 + 0.84\,t \,, 
\end{equation}
\begin{equation}\label{F22}
\alpha_{K^{*}_2} = 0.11 + 0.93\,t \,. \vspace*{0.3 \baselineskip}
\end{equation}
The errors in the coefficients in (\ref{F21}), (\ref{F22}) do
not exceed 2\%, which is insignificant for our purposes.

\begin{figure*}
 \includegraphics[width=0.31\textwidth]{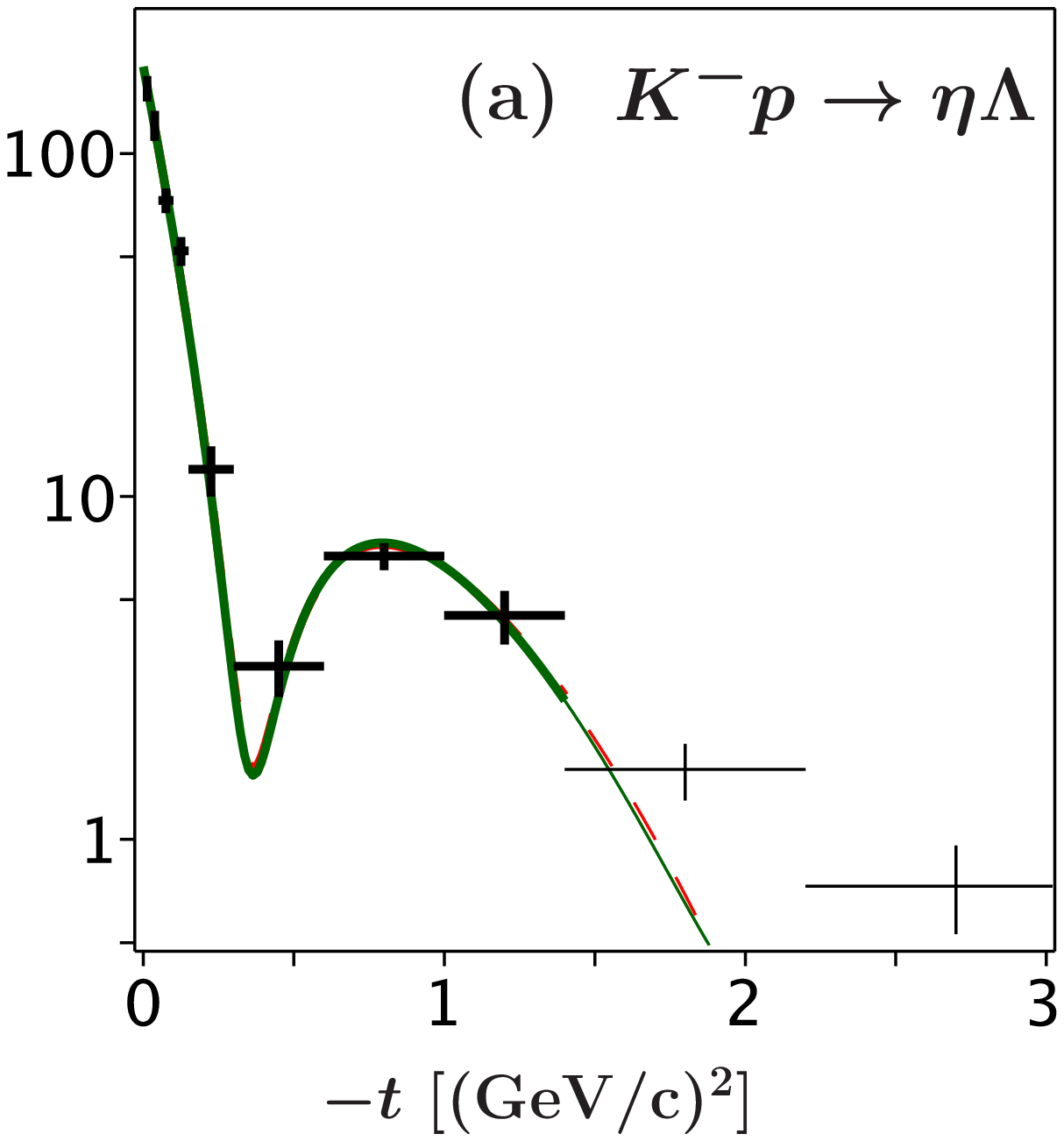}
 \hfill 
 \includegraphics[width=0.31\textwidth]{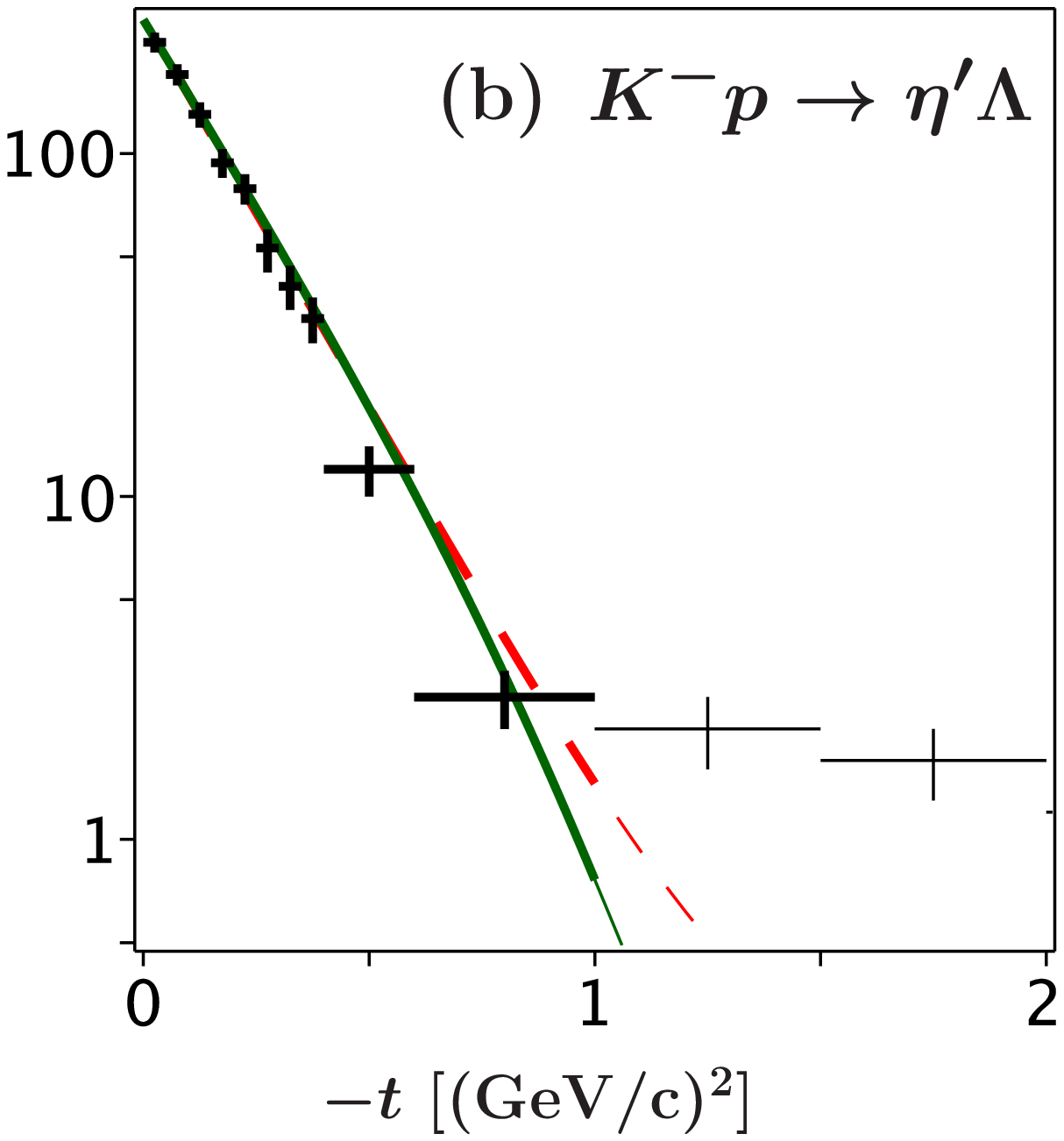}
 \hfill 
 \includegraphics[width=0.31\textwidth]{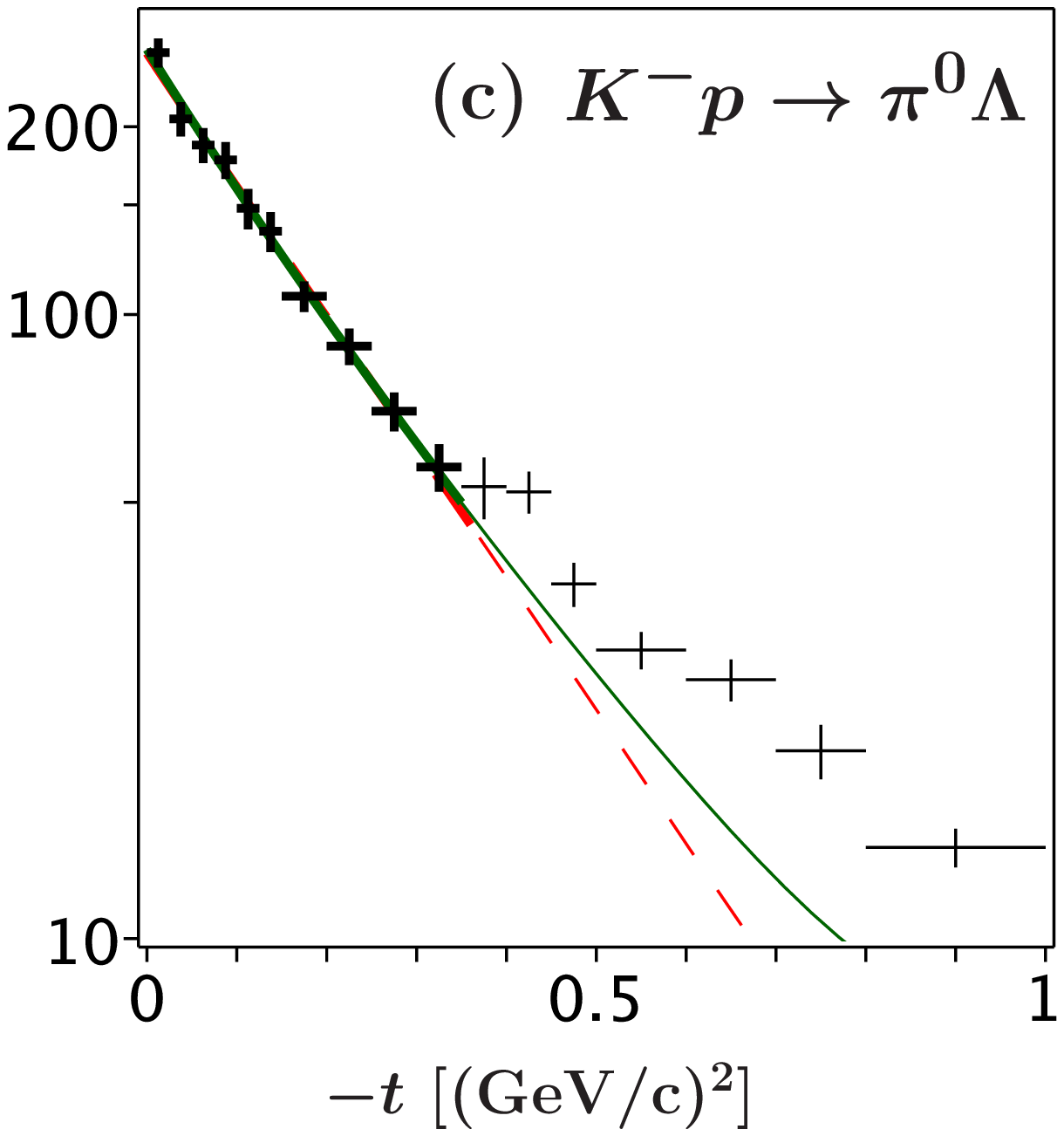}
\caption{\small Differential cross-sections $K^{-} p \to (\eta, \eta',
\pi^{0}) \Lambda$ [$\mu$b/(GeV$\!$/c)$^2$] at 4.2 GeV$\!$/c
\cite{Marzano}. Theoretical curves in (a), (b), (c) correspond to
(\ref{F13}), (\ref{F14}), (\ref{F18}), respectively. Solid (green) and
dashed (red) curves represent solutions Fit1 and Fit2, respectively. In
the regions of the fit the data and curves are shown by thick lines.}
\label{Fig2} \vspace*{\baselineskip}
\end{figure*}

Further, at relatively low energies we use formulas (\ref{F13}), 
(\ref{F14}) and (\ref{F18}). The differential cross-sections in this
energy range were measured by several groups
\cite{Moscoso,Mason,Marzano}, but only data \cite{Marzano} at 4.2
GeV$\!$/c are suitable for the fit. Unfortunately, we can not use the
totality of these data as they cover a large area of the transfer while
our formulas with one-reggeon exchanges are valid at small $t$ only.
Since we do not {\it a priori} know how small $t$ should be, we do a
series of the fits gradually expanding the area of $t$. We continue this
procedure until the quality of the description worsens sharply or
physical solutions disappear. In this manner we define the limiting
values $-t =$ 1.4, 1.0, 0.35 (GeV$\!$/c)$^2$ with the numbers of
experimental points 8, 10, 10 in the cases with $\eta$, $\eta'$,
$\pi^0$, respectively. In the mentioned areas there is only one
solution. We present it in table~\ref{T1} as Fit1. (The contributions to
$\chi^2$ associated with $\eta$, $\eta'$, $\pi^0$ are 4.5, 3.8, 3.2,
respectively.) In fig.~\ref{Fig2} we show this solution by solid (green)
thick curves. The solid~thin curves continue the solution beyond the
region of the fit. The discrepancy with the data in this region means
that some unaccounted contributions dominate. Typically, they are
reggeon-pomeron or multi-reggeon contributions characterized by a lower
slope and hence prevailing at large $-t$ \cite{Regge}. 

As we can see, the mentioned solution describes data well, but most
parameters are determined with large errors. In particular, the
mixing angle $\theta$ in essence is undetermined. So we can assign a
certain value to $\theta$ and then with this value do the fit again. We
take $\theta = -20.8^{^{\mbox{\scriptsize o}}}$, the value obtained
below. In this case again there is only one solution. We present it in
table~\ref{T1} as Fit2. (The contributions to $\chi^2$ associated with
$\eta$, $\eta'$, $\pi^0$ are 4.4, 4.0, 3.7, respectively.) In
fig.~\ref{Fig2} we show it by dashed (red) curves. We see that Fit1 and
Fit2 coincide within the errors, and the curves almost coincide in the
region of the fit. This means that in terms of conformity to the data
both solutions are equivalent.  

\begin{table*}
\caption{Solutions of the fit of data \cite{Marzano} (Fit1, Fit2) and
\cite{GAMS} (Fit3--Fit6). Fit1--Fit3 and Fit4--Fit6 match the coherent
and non-coherent mode, respectively. Parameters $g_{_{T}}$, $g_{_{V}}$
are given in $\mu$b$^{1/2}$/(GeV$\!$/c), the slopes in
(GeV$\!$/c)$^{-2}$.
}\label{T1}
\begin{tabular*}{\textwidth}{@{\extracolsep{\fill}}lllllll@{}}
\hline\noalign{\medskip}
            & Fit1    & Fit2    & Fit3   & Fit4    & Fit5    & Fit6 
\\[1mm]
\hline\noalign{\medskip}
{\footnotesize  $\chi^2/\mbox{d.o.f.}$ } $\!\!\!\!\!\!\!$
            & 11.4/19 & 12.1/20 & 36.4/37 & 37.4/37 & 35.0/37 & 37.5/39
\\[1mm]
{\footnotesize  $\theta$ [$\, ^{^{\mbox{\scriptsize o}}} \,$] }
                 & {\footnotesize $\!\!\!\!\!-12   \pm 12$  }
                 & {\footnotesize $\!\!\!\!\!-20.8$ [input] }
                 & {\footnotesize $\!\!\!\!\!-25.9 \pm 2.32$}
                 & {\footnotesize $\!\!\!\!\!-21.8 \pm 5.0$ }
                 & {\footnotesize $\!\!\!\!\!-22.2 \pm 4.3$ }  
                 & {\footnotesize $\!\!\!\!\!-20.8 \pm 4.9$ }
\\[1mm]
{\footnotesize  $g_{_{T}}$ }     
                 & {\footnotesize $8.1  \pm 5.3$ }
                 & {\footnotesize $12.2 \pm 4.4$ }
                 & {\footnotesize $1.7  \pm 1.3$ }
                 & {\footnotesize $2.0  \pm 0.4$ }
                 & {\footnotesize $2.1  \pm 0.1$ }
                 & {\footnotesize $2.07 \pm 0.09$}
\\[1mm]
{\footnotesize  $c_{_{T}}$  }
                 & {\footnotesize $1.2  \pm 0.8$ }
                 & {\footnotesize $1.5  \pm 0.8$ }
                 & {\footnotesize $12.1 \pm 6.4$ }
                 & {\footnotesize $4.1  \pm 0.4$ }
                 & {\footnotesize $4.3  \pm 0.3$ }
                 & {\footnotesize $4.1  \pm 0.2$ }
\\[1mm]
{\footnotesize  $g_{_{V}}$  }    
                 & {\footnotesize $12.2 \pm 0.6$ } 
                 & {\footnotesize $12.8 \pm 0.6$   }
                 & {\footnotesize $2.2  \pm 0.4$ }
                 & {\footnotesize $0.2  \pm 1.4$  }
                 & {\footnotesize $\!\!\!\!\!-2.2 \pm 1.2$ }
                 & {\footnotesize $0$ [input]    }
\\[1mm]
{\footnotesize  $c_{_{V}}$  }    
                 & {\footnotesize $1.7 \pm 0.1$ }
                 & {\footnotesize $1.7 \pm 0.1$ }
                 & {\footnotesize $3.0  \pm 0.9$ }
                 & {\footnotesize $0.7  \pm 13.7$}
                 & {\footnotesize $110  \pm 70$  }
                 & ---
\\[1mm]
{\footnotesize  $\xi_0$ }         
                 & {\footnotesize $1.13 \pm 0.77$ }
                 & {\footnotesize $0.72 \pm 0.31$ }
                 & {\footnotesize $1.02 \pm 0.86$ }
                 & {\footnotesize $0.96 \pm 0.18$ }
                 & {\footnotesize $0.91 \pm 0.08$ }
                 & {\footnotesize $0.95 \pm 0.08$ }
\\[1mm]
{\footnotesize  $\xi'$  }        
                 & {\footnotesize $1.0 \pm 0.8$ } 
                 & {\footnotesize $0.8 \pm 0.9$ }
                 & {\footnotesize $\!\!\!\!\!-10.0 \pm 6.5$ }
                 & {\footnotesize $\!\!\!\!\!-2.3 \pm 0.6$ }
                 & {\footnotesize $\!\!\!\!\!-2.4 \pm 0.4$ }
                 & {\footnotesize $\!\!\!\!\!-2.2 \pm 0.4$ }
\\[1mm]
{\footnotesize  $\zeta_0$ }
                 & {\footnotesize $1.19 \pm 0.07$ }
                 & {\footnotesize $0.91 \pm 0.24$   }
                 & --- & --- & --- & --- 
\\[1mm]
{\footnotesize  $\zeta'$  } 
                 & {\footnotesize $\!\!\!\!\!-0.6 \pm 0.5$ }
                 & {\footnotesize $\!\!\!\!\!-0.5 \pm 0.4$ }
                 & --- & --- & --- & --- 
\\[1mm]
\noalign{\smallskip}\hline
\vspace*{\baselineskip}
\end{tabular*} 
\end{table*}

We can clarify the reason why angle $\theta$ in Fit1 is poorly defined.
The point is that formula (\ref{F18}) for the cross-section with $\pi^0$
does not contain $\theta$, while dependence on $\theta$ in formulas
(\ref{F13}), (\ref{F14}) in the tensor channel appears in the
combination $\theta - \widetilde\delta$. So when considering the two
latter reactions, we get a situation which is similar to that in the
case in the $\pi^-$ beams, although with a correction for the presence
of two trajectories. Namely, two series of solutions with close $\chi^2$
appear, which differ by a sign of the interference between contributions
of the trajectories. All these solutions are distributed in the
($\theta,\,\xi_0$)-plane along the curve $\theta - \widetilde\delta_0 =
\mbox{constant}$, where $\widetilde\delta_0 = \widetilde\delta (0)$.
Simultaneously, they all describe the dip in $K^{-} p \to \eta \Lambda$
and monotonic behavior of $K^{-} p \to \eta' \Lambda$. The inclusion of
data with $\pi^0$ excludes one of the series with ``wrong''
interference, and fixes $\theta$, $\xi_0$ via the correlation with other
parameters. However, fixing via the correlation implies poor definition,
as it happens.
 
\begin{figure*}
 \includegraphics[width=0.31\textwidth]{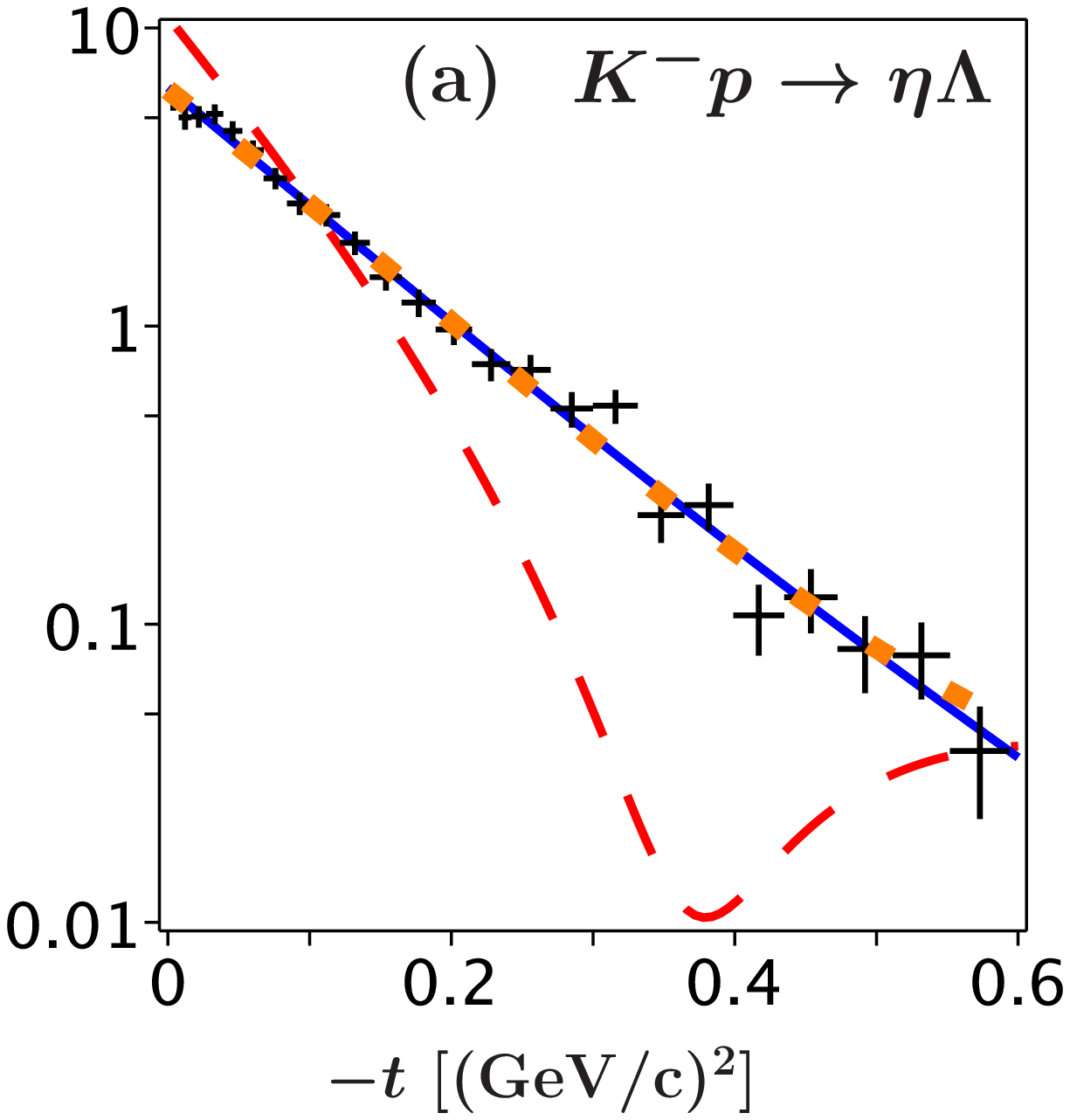}
 \hfill 
 \includegraphics[width=0.31\textwidth]{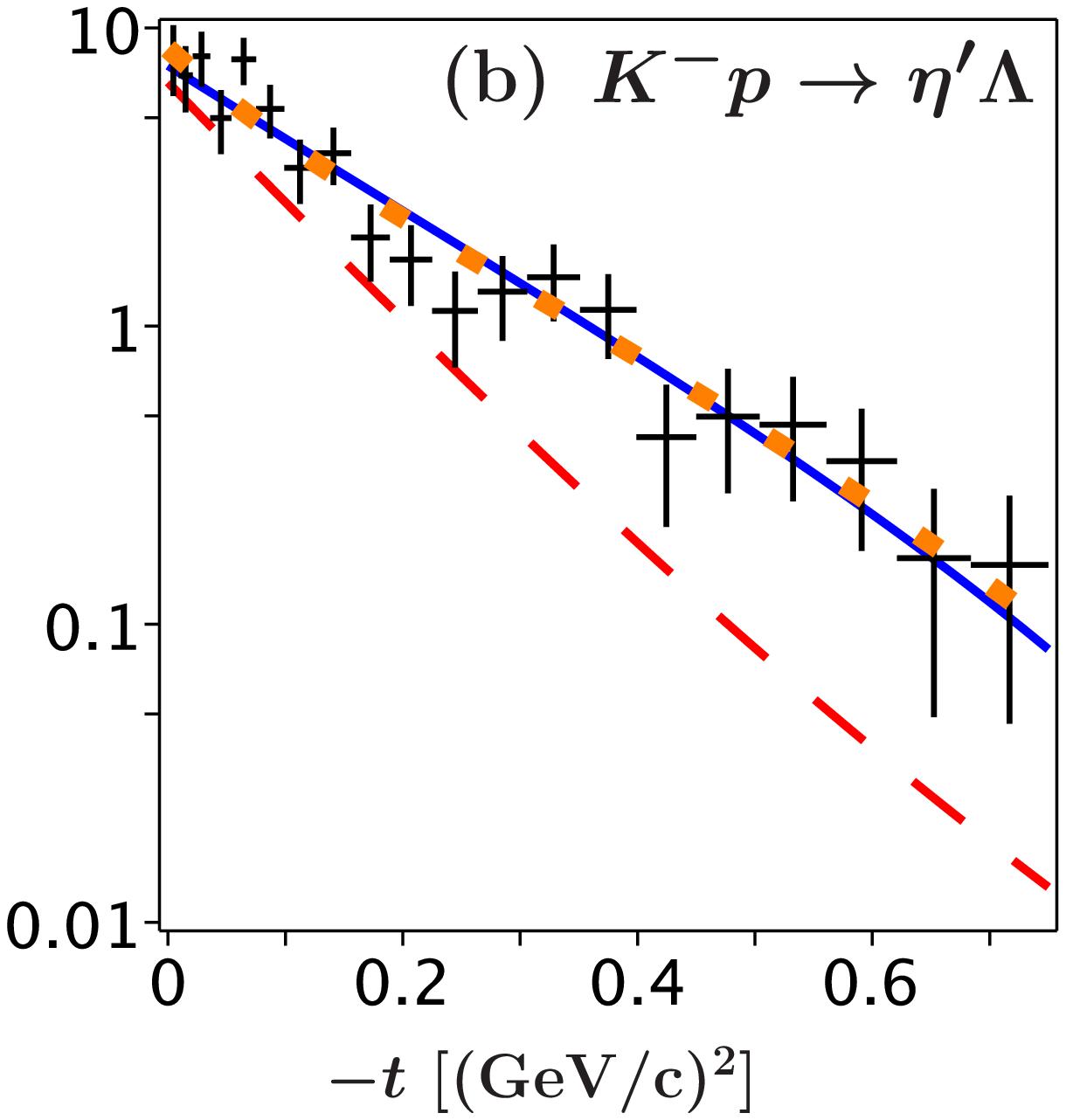}
 \hfill 
 \includegraphics[width=0.31\textwidth]{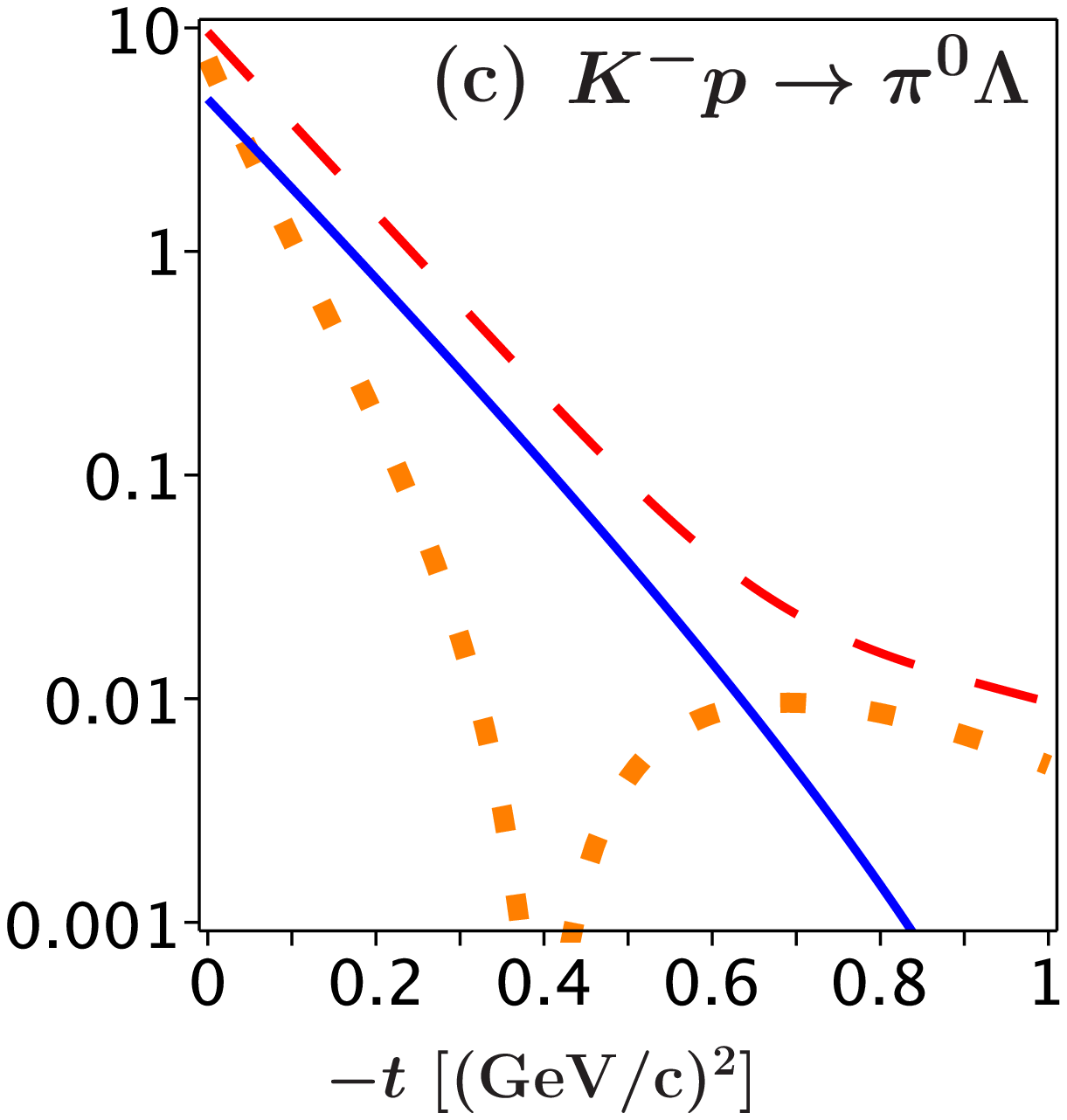}
 \hfill 
\caption{\small Differential cross-sections $K^{-} p \to (\eta, \eta',
\pi^{0}) \Lambda$ [$\mu$b/(GeV$\!$/c)$^2$] at 32.5 GeV$\!$/c. The data
in (a) and (b) are taken from \cite{GAMS}. Solid (blue) curves represent
Fit6. Solution Fit2 in recalculation to the energy \cite{GAMS} is shown
by (red) dashed curves. Solution Fit3 is shown by (orange) points. In
the frame (c) the curves for Fit3 and Fit6 are shown at $\zeta =1$ in
(\ref{F18})} \label{Fig3}  
\end{figure*}

Now we turn to relatively high energies. The available data in this
range are \cite{GAMS}, obtained at the $K^-$ momentum 32.5 GeV$\!$/c.
Unfortunately, these data cover reactions with $\eta$ and $\eta'$ only.
However, the fit in this case leads to more precise outcomes. 

First of all, we check whether there is a solution in the coherent mode.
It turns out that such a solution exists, see Fit3 in table~\ref{T1}.
Its essential features are the negative sign of $\xi'$ and the
approximate equality in absolute value of $\xi'$ and $c_{_{T}}$. The
negative sign of $\xi'$ means exponential growth of $\xi(t)$ with
increasing $-t$. So at $t = - 0.4$ (GeV$\!$/c)$^2$ we have $\xi \approx
55$. Such a great value destroys the mechanism of appearance of a dip in
$K^{-} p \to \eta \Lambda$. However, both features are unacceptable to a
physical solution since they mean large splitting between the slope
parameters in the vertex function. Really, the slope in the pure singlet
component in this case is $c_{_{T}} + \xi' = 2.1 \pm 0.3$ (strong
correlation is taken into account). This is almost 6 times smaller than
the slope $c_{_{T}}$ in the octet component. The difference is too great
for the flavor symmetry violation. So Fit3 must be considered as
unphysical solution. However that is not all. Besides the slopes in Fit3
are incompatible with Regge behavior. Really, $\xi'$ should be unchanged
with changing the energy while $c_{_{T}}$ and $c_{_{V}}$ should evolve
in accordance with (\ref{F4}). However, $\xi'$ in Fit3 changes the sign
and in absolute value exceeds tenfold $\xi'$ in Fit1. Moreover,
proceeding from Fit2, $c_{_{T}}$ and $c_{_{V}}$ both must equal 3.3 at
the energy of Fit3. However, $c_{_{T}}$ in Fit3 is significantly larger.
This means violation of Regge behavior. The mentioned inconsistencies
are clearly visible in fig.~\ref{Fig3}(a,b), where solution Fit2
recalculated to the energy of Fit3 is shown by dashed (red) curves, and
solution Fit3 is shown by (orange) points. In fig.~\ref{Fig3}(c) the
analogous curves have meaning of predictions for $K^{-} p \to \pi^0
\Lambda$ at 32.5 GeV$\!$/c. 

On the basis of the above discussion, we conclude that the high-energy
data are incompatible with predictions of the Regge approach. So in this
energy range a modified approach is required. For this purpose we take
advantage of our model in the non-coherent mode. 

The corresponding fit of data \cite{GAMS} with formulas (\ref{F16}),
(\ref{F17}) leads to two solutions, shown in table~\ref{T1} as Fit4 and
Fit5. They almost coincide, except $g_{_{V}}$ and $c_{_{V}}$. Namely,
$g_{_{V}}$ in Fit4 is compatible with zero, whereas $g_{_{V}}$ in Fit5
is not. Besides, $c_{_{V}}$ in Fit5 is anomalously large. However, on
close examination the mentioned differences are illusory. Really, due to
the large $c_{_{V}}$ the vector contribution in Fit5 is nonzero at $t
\approx 0$ only. So, both solutions are compatible with zero
contribution of the vector trajectory. Assuming initially
$g_{_{V}}\!=\!0$, we come to solution Fit6. In view of above discussion,
Fit6 is practically equivalent to Fit4 and Fit5. In fig.~\ref{Fig3} we
show Fit6 by solid (blue) curves.

An essential feature of the solutions Fit4-Fit6 as compared to Fit1, are
essentially smaller errors in $\theta$ and $\xi_0$. This is a
consequence of the fact that both these parameters are determined based
on the data in the reactions with $\eta$ and $\eta'$ only, without
involving the data with $\pi^0$ (see above discussion). As the final
outcome for $\theta$ we take its value in Fit6,
\begin{equation}\label{F23}
  \theta = -(20.8 \pm 4.9)^{\mbox{\scriptsize o}}.
\end{equation} 

Our main conclusion in this subsection is that  in the framework of
conventional Regge approach the data in the $K^{-}$ beams are well
described at relatively low energies, but are no longer described at the
transition to higher energies. At the same time, at the higher energies
the data are well described in the mode with non-coherent summation of
intermediate contributions. The data fit at low energies does not
determine the $\eta$--$\eta'$ mixing angle. The fit at higher energies
leads to estimate (\ref{F23}).

\subsection{Gluonium admixture}\label{sec3.3}

The above analysis was carried out in the particular scheme (\ref{F10})
of the $\eta$--$\eta'$ mixing. However analysis can easily be
generalized to any scheme. Here we consider a generalization which
includes a gluonium admixture in the $\eta'$. In this scheme
\begin{eqnarray}\label{F24}
|\eta \rangle \; &=&
 \cos\theta \, |\eta^8 \rangle -  \sin\theta \, |\eta^0 \rangle\,,
 \nonumber\\[0.4\baselineskip]
 |\eta' \rangle &=&
 \cos\theta_G \Bigl(\sin\theta \, |\eta^8 \rangle +
 \cos\theta \, |\eta^0 \rangle \Bigr)\! +
 \sin\theta_G \, |\eta^G \rangle. \qquad\quad
\end{eqnarray}
where $|\eta^G \rangle$ is a gluonium state and $\theta_G$ is an
additional mixing angle. This scheme was first proposed as a solution to
the axial Ward identities for the relevant composite interpolating
fields on the condition of the renormalization-group invariance of the
pattern of the mixing \cite{N2}. Afterwards this scheme was repeatedly
considered on purely phenomenological basis, see bibliography in
\cite{N1}. In scheme (\ref{F24}) the above formulas do not change in the
case of $\eta$ and only slightly change in the case of $\eta'$. Namely,
a factor $\cos^2 \theta_G$ appears in the r.h.s. in (\ref{F8}),
(\ref{F11}), (\ref{F14}), (\ref{F17}). We complement the allowable
range of parameters by condition $0.5 < \cos^2 \theta_G \le 1$, where
the lower bound means that $\eta'$ is predominantly a quark state, not a
glueball.

The analysis of data \cite{GAMS} in the $\pi^-$ beams, similar to that
in sect.~\ref{sec3.1}, leads to a series of solutions with close
$\chi^2$ grouped in the ($\theta,\xi_0,\theta_G$)-space near a surface
specified by condition
\begin{equation}\label{F25}
R_{\pi}^{\,\eta'/\eta}(0) \> = \>
\cos^2 \theta_G \; \tan^2 (\theta + \theta_{id} - \delta_0) \,.
\end{equation}
Among these solutions $\cos^2 \theta_G$ varies from $0.68 \pm 0.14$ to
$0.92 \pm 0.20$, in all cases with $\chi^2/\mbox{d.o.f.} \approx 30/34$.
Thereby, the data in the $\pi^-$ beams do not provide solid information
about the gluonium admixture in $\eta'$.

The fit of data \cite{Marzano} and \cite{GAMS} in the $K^-$ beams with
all the parameters free, do not reveal any physical solutions.
Nevertheless, at various fixed $\cos^2 \theta_G$ the solutions appear,
and we can trace their evolution. Recall that at $\cos^2 \theta_G = 1$
all the solutions are presented in table~\ref{T1}. With decreasing
$\cos^2 \theta_G$ from 1, the $\chi^2$ is typically increasing in the
appropriate branches of the solutions with the parameters are smoothly
evolving. In so doing, the solutions in the branches Fit1 and Fit3
continue to be incompatible with each other, and the solutions in Fit3
are characterized by abnormal splitting of the slope parameters as was
discussed in sect.~\ref{sec3.2}. 

Within $0.6 \le \cos^2 \theta_G \le 1$ the solutions exist in all cases
with free $\theta$. In particular, at $\cos^2 \theta_G = 0.6$ the
$\chi^2/\mbox{d.o.f.}$ in branches Fit1 and Fit6 takes values 11.7/19
and 38.0/39, respectively, and in both cases $\theta$ remains within the
errors of table~\ref{T1}. In addition, in the branch Fit2 with $\theta =
-20.8^{\mbox{\scriptsize o}}$ and free $\cos^2 \theta_G$, there is only
one solution:
$\cos^2 \theta_G = 0.68 \pm 0.67$, $g_{_{T}} = 14.3 \pm 7.9$, $c_{_{T}}
= 1.6 \pm 0.8$, $g_{_{V}} = 12.8 \pm 0.6$, $c_{_{V}} = 1.7 \pm 0.1$,
$\xi_0 = 0.74 \pm 0.30$, $\xi' = 0.7 \pm 1.0$, $\zeta_0 = 0.82 \pm
0.33$, $\zeta' = - 0.5 \pm 0.5$, $\chi^2/\mbox{d.o.f.} = 12.0/19$.
Notice that $\cos^2 \theta_G$ is compatible with 1.

So the data on the charge-exchange reactions are rather indifferent
to a possible gluonium admixture in the $\eta'$, and on this basis we
can not draw a conclusion about the presence or absence of this
admixture. However the conclusion about the mode change of summation of
intermediate contributions remains in force in the presence of the
gluonium admixture. In the case of small gluonium admixture the estimate
for the angle $\theta$ undergoes insignificant changes.

\section{Solution at intermediate energies}\label{sec4}

We have seen above that the charge-exchange reactions in the $K^{-}$
beams at relatively low and high energies are well described by the
formulas obtained in the coherent and non-coherent modes, respectively.
Now we consider the problem of the description in the intermediate
energy region where contributions of both modes are possible.

We offer a solution in the spirit of the density matrix. Specifically,
we assume that each mode can be implemented with a certain probability.
Since there is no third, the sum of the two probabilities must be equal
to one. So, we consider the differential cross-section in the~form
\begin{equation}\label{F26}
\frac{{\rm d} \sigma}{{\rm d} t} (s,t) \;=\;
w \; \frac{{\rm d} \sigma_{\mbox{\scriptsize c}}}{{\rm d} t} (s,t) \;+\;
(1-w) \; \frac{{\rm d} \sigma_{\mbox{\scriptsize nc}}}{{\rm d} t}
(s,t)\,.
\end{equation} 
Here $w$ is the probability to find the system in the coherent mode, 
indices ``c'' and ``nc'' mean the modes of coherent and non-coherent
summation. In the most general case $w$ is a function of $s$ and $t$.
Asymptotically at low and high $s$, we expect $w \to 1$ and $w \to 0$,
respectively. Our task is to find an algorithm of experimental
measurement of $w$.

Below, we do this under the assumption that at small transfer one can
neglect in $w$ the dependence on $t$, so that $w = w(s)$. In fact,
reliable measurements are needed at least at two energies. At $s=s_1$ we
have
\begin{equation}\label{F27}
\frac{{\rm d} \sigma}{{\rm d} t} (s_1,t) = \left[
w_1  \frac{{\rm d} \sigma_{\mbox{\scriptsize c}}}{{\rm d} t}
(s_1,t)\right] +
\left[ (1 \! - \! w_1 ) \frac{{\rm d} 
\sigma_{\mbox{\scriptsize nc}}}{{\rm d} t} (s_1,t)\right] \! . \;
\end{equation} 
Here $w_1 = w(s_1)$, ${\rm d} \sigma_{\mbox{\scriptsize c}} / {\rm d} t$
and ${\rm d} \sigma_{\mbox{\scriptsize nc}} / {\rm d} t$ are described
by appropriate formulas in the coherent and non-coherent modes. The
large square brackets mean that factors $w_1$ and $(1-w_1 )$ are not
considered as independent parameters, but at the fit are considered
absorbed by normalization constants in the cross-sections.
At $s = s_2$ an analogous formula includes $w_2 = w(s_2)$ as a hidden
parameter. However, we write this formula in a modified form. Namely,
we factor out numbers $x$ and $y$, defined as 
\begin{equation}\label{F28}
x = \frac{w_2}{w_1}\,, \qquad \quad
y = \frac{1-w_2}{1-w_1}\,.
\end{equation} 
So at $s=s_2$, we have
\begin{equation}\label{F29}
\displaystyle
\frac{{\rm d} \sigma}{{\rm d} t} (s_2,t) = x \left[ 
w_1 \frac{{\rm d} \sigma_{\mbox{\scriptsize c}}}{{\rm d} t} (s_2,t)
\right]
 + y \left[(1 \! -\! w_1 ) \frac{{\rm d} 
\sigma_{\mbox{\scriptsize nc}}}{{\rm d} t} (s_2,t) \right] \! . \quad
\end{equation} 

Further, we notice that all the parameters in the square brackets in
(\ref{F29}) are actually determined by the fit at $s=s_1$ with
(\ref{F27}). Really, some parameters are unchanged at changing the
energy, and the remaining ones are changed by means of (\ref{F4}) with
$\tilde{s}_0 = s_2$, $s_0 = s_1$. So in (\ref{F29}) only numerical
factors $x$ and $y$ are unknown. They may be determined based on the fit
at $s=s_2$. Then, solving system (\ref{F28}), we obtain the
probabilities:
\begin{equation}\label{F30}
w_1  =      \frac{y-1}{y-x}\,, \qquad \quad
w_2 = x \, \frac{y-1}{y-x}\,.
\end{equation}

In this way, we have presented an algorithm to measure $w(s)$. Of
course, in practice it is expedient to do a joint fit of data
simultaneously at different energies. On this basis we can determine
$w(s)$ in a wide range of energy. We expect that $w(s)$ is monotonically
decreasing with increasing the energy from asymptotic value 1 at
relatively low energies up to asymptotic value 0 at higher energies.

Unfortunately, the currently available data are not sufficient to carry
out such an analysis because data \cite{Marzano} contain too few
measured points and \cite{GAMS} cover too small region of $t$. Besides,
in the case of the common use the data \cite{GAMS} from the very
beginning must be recalculated to units of the differential
cross-sections, which leads to an additional 12\%-error, see appendix A.
This further reduces the quality of the fit. As for the data
\cite{Harran} at 8.25 GeV$\!$/c, which fall into the intermediate
region, their quality is not sufficient to do the fit with them because,
in particular, they contain too few measured points.

\begin{figure*}
\hspace*{0.08\textwidth}
\includegraphics[width=0.31\textwidth]{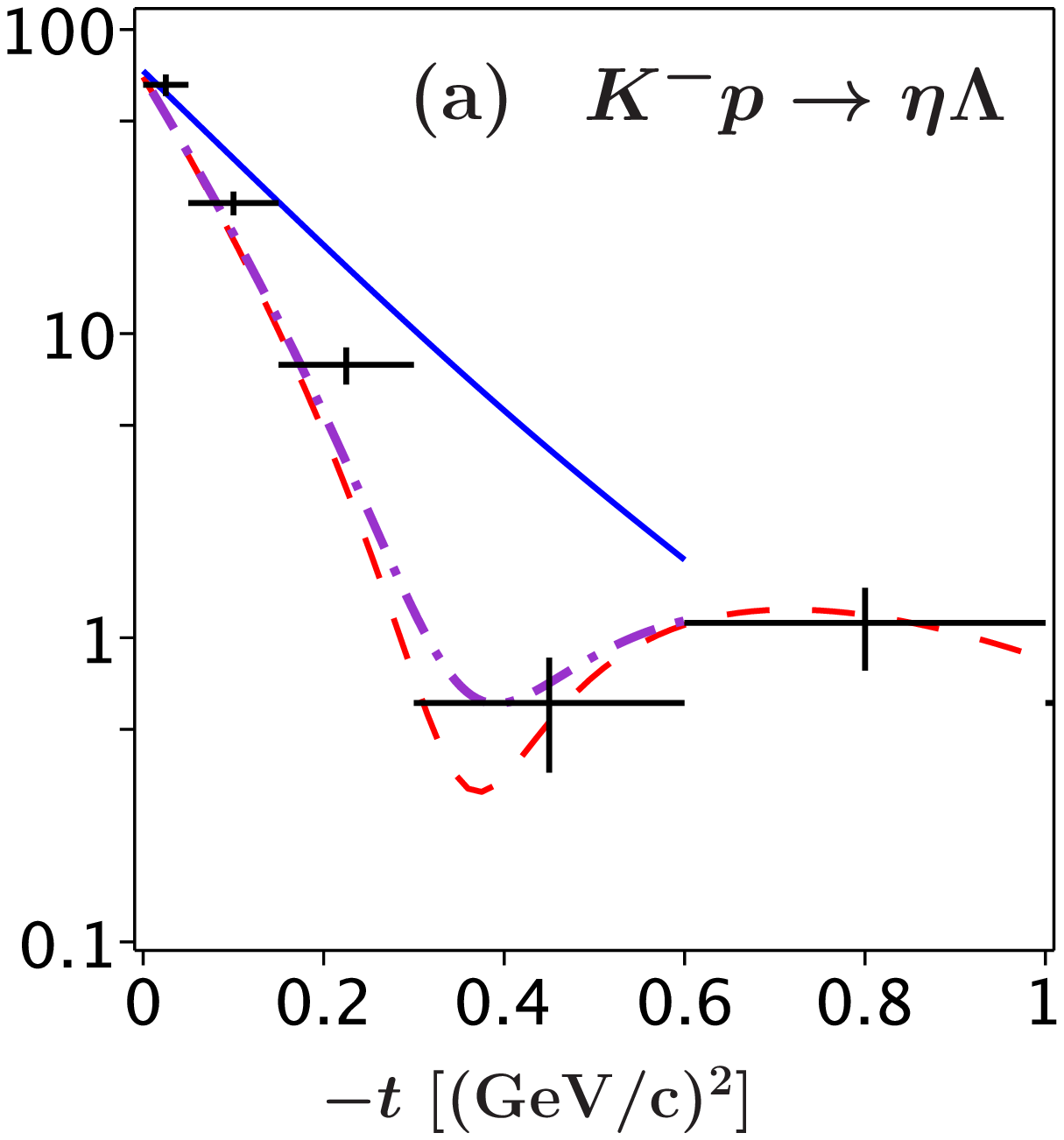}
 \hfill 
 \includegraphics[width=0.31\textwidth]{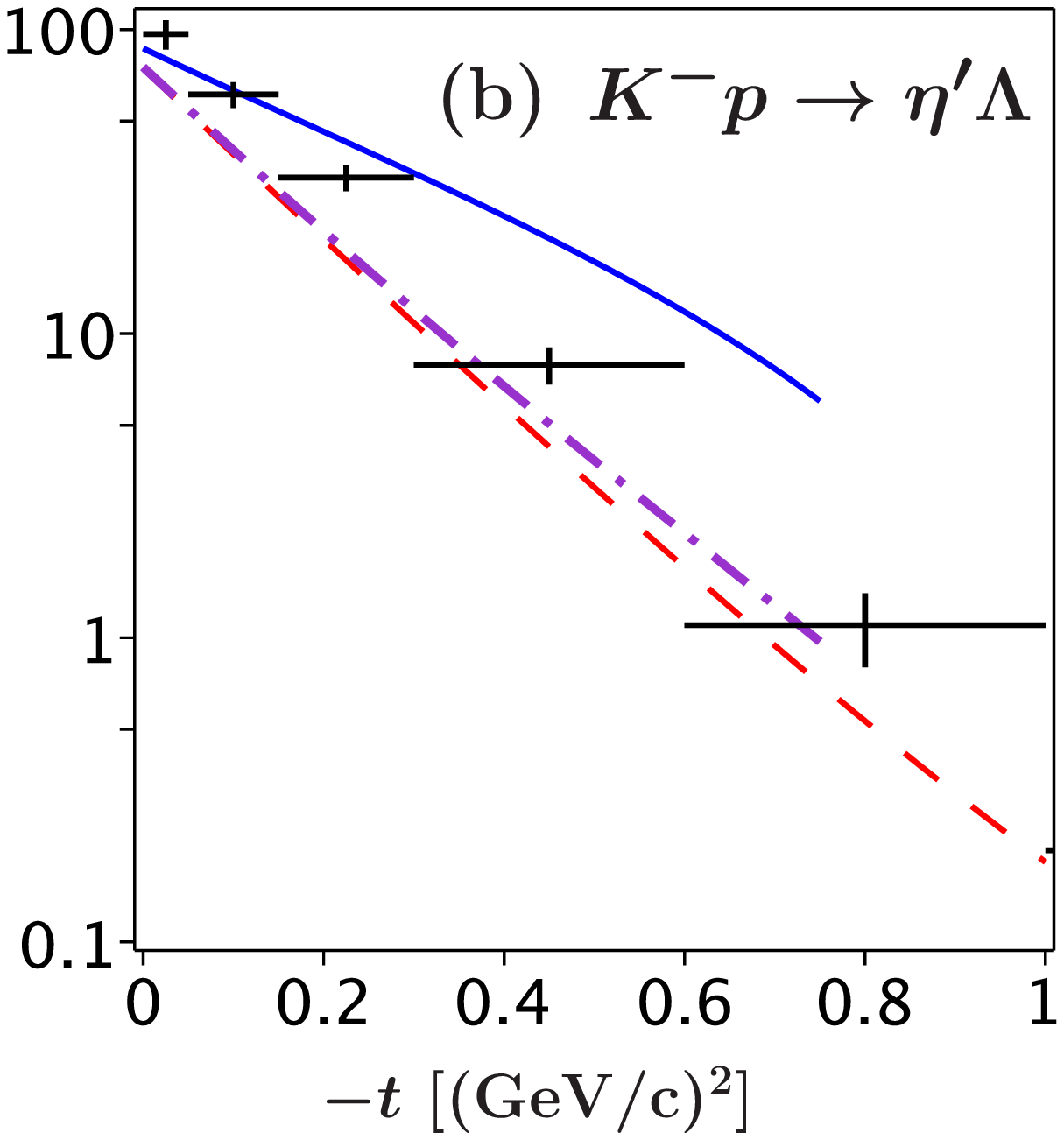}
\hspace*{0.1\textwidth}
\caption{\small Differential cross-sections $K^{-} p \to (\eta, \eta')
\Lambda$ [$\mu$b/(GeV$\!$/c)$^2$] at 8.25 GeV$\!$/c \cite{Harran}. Solid
(blue) and dashed (red) curves represent solutions Fit6 and Fit2,
respectively, recalculated to energy \cite{Harran}. Dash-dot (violet)
curves represent solution built by (\ref{F26}) at $w=0.95$.}\label{Fig4}
\end{figure*}

However, all the mentioned data may be used for illustrative purposes.
Namely, let us suppose that at the energies of \cite{Marzano} and
\cite{GAMS} the probability $w$ takes values 1 and 0, respectively.
Then, recalculating solutions Fit2 and Fit6 to the intermediate energy
\cite{Harran} and substituting the result into (\ref{F26}), we obtain a
prediction for the differential cross-section at 8.25 GeV$\!/c$. In
fig.~\ref{Fig4} we show data \cite{Harran} and we present by dashed
(red) and solid (blue) curves solutions Fit2 and Fit6, respectively,
recalculated to this energy. The dash-dot curve gives the result defined
by (\ref{F26}) at $w=0.95$, where 0.95 is chosen arbitrarily. We see
that the latter curve lies between the former two and on the whole
better matches the data. One should bear in mind, however, that the
positions of the curves in fig.~\ref{Fig4} are shown without taking into
account the appropriate errors.

\section{Discussion and conclusion}\label{sec5}

The analysis in this work confirms the earlier qualitative conclusion
\cite{N1} about the mode change with increasing the energy of summation
of intermediate contributions in the charge-exchange reactions. In
support of this conclusion, we put forward the following results
obtained based on the fit of data. First, we found a solution which
describes in the Regge approach the charge-exchange reactions $K^{-} p
\to (\eta, \eta', \pi^{0}) \Lambda$ at relatively low energies. In our
model this solution implies the mode of coherent summation of
intermediate contributions. Then, we showed the absence of a similar
solution at relatively high energies. Instead, a solution appears that
corresponds to the mode of non-coherent summation. Finally, we compared
the obtained solutions by recalculating to a common energy and
established fundamental difference between them. 

We emphasize that the above solutions are statistically well-founded.
Moreover, the data on the charge-exchange reactions  in the $\pi^-$
beams are fitted very well.\footnote{In the case of $\pi^-$ beams the
mode change is independently confirmed by effect of the energy
dependence of the slope of $R_{\pi}^{\,\eta'/\eta}(t)$, see discussion
in sect.~\ref{sec3.1} and in \cite{N1}.} This indicates that our model
describes the charge-exchange reactions well, and results obtained on
its basis are reliable. 

At the same time, the range of applicability of the model is limited by
small transfers. This is a consequence of the fact that the model takes
into account the contributions of leading trajectories only,\footnote{We
do not take into consideration the daughter trajectories as they have
close slope but lower intercept. Therefore their contributions are
suppressed.} and ignores the contributions of the Regge cuts that manifest
themselves at higher $-t$. In our approach the region where the
contributions of the cuts may be ignored is determined empirically by
gradually increasing the upper limit of $|t|$ until the data are well
described. So, in principle, the range of applicability of the model can
be expanded by including in analysis the reggeon-pomeron or
multi-reggeon contributions. We do not consider such an extension since
this implies a substantial increasing in the number of parameters in
conditions of
limited data.

In our analysis we proceeded from the assumption that the angle of the
$\eta$--$\eta'$ mixing is unknown, and we considered it as a parameter
which is to be determined by the fit. This approach is valuable in
itself in view of the lack of well-established estimate of the mixing
on the basis of the charge-exchange reactions \cite{Gilman,Scadron}.
Furthermore, unlike \cite{N1} we considered the parameter of violation
of the nonet symmetry as unknown, which must be determined by the fit,
as well. Unfortunately, this leads to the impossibility of determining
the mixing angle on the basis of data in the $\pi^-$ beams, since
corresponding cross-sections are determined by the difference of the
mixing angle and the parameter of the nonet symmetry breaking. For this
reason, we determined the mixing by basing on the data in the $K^-$
beams. Unluckily, the appropriate estimate includes large error and
therefore in practical terms is little informative. However, we show
that the gluonium admixture in the $\eta'$ has little effect on the $\eta
$--$ \eta'$ mixing. Moreover, the gluonium admixture does not
affect the conclusion about the mode change with increasing the energy.

Finally, we found a solution to the problem of description of the
charge-exchange reactions at the intermediate energy range where both
modes are possible, the coherent and non-coherent. The solution is based
on an idea that each mode occurs with a certain probability, and the
probability is dependent on the energy. By this means the differential
cross-section is determined as a sum of contributions with certain
modes, weighted with the probabilities of finding the system in the
particular mode. We propose an algorithm for measuring the
probabilities. A study of the charge-exchange reactions with taking into
account this solution might be a subject of future research provided
that sufficiently detailed data become available.

In summary, with high reliability we confirmed the effect of the mode
change with increasing the energy of summation of intermediate
contributions in the differential cross-sections of charge-exchange
reactions. The analysis in this paper more thoroughly specifies the
model for further systematic investigation of this phenomenon. In
particular, one can study the energy dependence of the coherent and
non-coherent contributions to the cross-sections of the mentioned
reactions. 

\bigskip\noindent
The author is grateful for useful discussions to V.V.Ezhela, V.A.Petrov,
R.N.Rogalyov, and S.R.Slabospitsky. Special thanks to V.D.Samoylenko for
providung the GAMS-4$\pi$ data and for pointing out on a possibility of
converting them to the form of differential cross-sections using the 
NICE results.

\medskip
\begin{flushleft}
{\bf\large Appendix A}
\end{flushleft}

The data \cite{GAMS} are presented in the form of numbers of pairs of
gamma-quanta arising due to the decays of $\eta$ and $\eta'$. Our task
is to determine the proportionality factor between the numbers of the
registered pairs of gamma-quanta and the corresponding differential
cross-sections.

In the case of $\pi^-$ beams we take advantage of formula (12) in
\cite{apel},
\begin{equation}\label{A1}
\left.\frac{\d\sigma}{\d t} (\pi^{-} p \to \eta  n \to \gamma\gamma n)
\right|_{t=0}\;=\;
(37.1 \pm 3.6) \left(\frac{s}{s_0}\right)^{-1.26 \pm 0.04} \times
\mu b \,(\mbox{GeV\!/c})^{-2} \,.
\end{equation}
Here $s_0 = 10\,(\mbox{GeV\!/c})^2$. At $s=62.1\,(\mbox{GeV\!/c})^2$,
which means $p_{\mbox{\scriptsize LAB}} = 32.5\,\mbox{GeV\!/c}$, the
r.h.s. in (\ref{A1}) is equal to $(3.71 \pm 0.45) \times \mu b
\,(\mbox{GeV\!/c})^{-2}$. On the other hand, at $p_{\mbox{\scriptsize
LAB}} = 32.5$ $\mbox{GeV\!/c}$ \cite{GAMS} obtained
\begin{equation}\label{A2}
\left.\frac{\d N}{\d t} (\pi^{-} p \to \eta  n \to \gamma\gamma n)
\right|_{t=0}\;=\;
(2.21 \pm 0.02) \times 10^{6} \times (\mbox{GeV\!/c})^{-2} \,,
\end{equation}
where $N$ is the number of pairs of gamma-quanta. The sought-for factor
is a ratio of (\ref{A1}) to (\ref{A2}). Numerically it is $r_{\pi^-} =
(1.68 \pm 0.20) \times 10^{-6} \mu b$. In the case of yield of $\eta'$
the appropriate factor numerically is the same, because the gamma-quanta
in \cite{GAMS} were registered in the same experiment. Moreover, this
factor determines also the ratio of ${\rm d} \sigma (\pi^{-} p \to
\eta^{[\prime]} n) / {\rm d} t$ to $\Delta N (\pi^-,\eta^{[\prime]}) /
\Delta t$, where $\Delta N (\pi^-,\eta^{[\prime]})$ is the number of
$\eta^{[\prime]}$ formed in the $\pi^-$ beam in the course of experiment
\cite{GAMS} in the bin $t$ of the width~$\Delta t$.

In the case of $K^-$ beams, we take advantage of the fact that the beam
of negative particles in \cite{GAMS} consisted of 98\% $\pi^-$ and 2\%
$K^-$, and the admixture of other particles was negligible. On this
basis, we get $r_{K^-} = (0.98/0.02) r_{\pi^-} = (8.23 \pm 0.98)
\times 10^{-5} \mu b$. Here $r_{K^-}$ has the meaning of the ratio of
${\rm d} \sigma (K^{-} p \to \eta^{[\prime]} n) / {\rm d} t$ to $\Delta
N(K^-,\eta^{[\prime]}) / \Delta t$, where $\Delta
N(K^-,\eta^{[\prime]})$ is the number of  $\eta^{[\prime]}$ formed in
the $K^-$ beam in the corresponding bin $t$ of the width $\Delta t$.

Given the branchings of $\eta^{[\prime]} \to \gamma\gamma$, the obtained
factors determine different conversion factors depending on the choice
of the final state.

\end{document}